\documentclass[twocolumn,aps,pra,showpacs]{revtex4}
\usepackage{amsmath}

\begin{document}

\title{Three-particle entanglement versus three-particle nonlocality}
\author{Jos\'{e} L. Cereceda}
\email{jl.cereceda@teleline.es}
\affiliation{C/Alto del Le\'{o}n 8, 4A, 28038 Madrid, Spain}
\date{published 12 August 2002}


\begin{abstract}
The notions of three-particle entanglement and three-particle nonlocality are discussed in the light of Svetlichny's inequality [Phys. Rev. D \textbf{35}, 3066 (1987)]. It is shown that there exist sets of measurements which can be used to prove three-particle entanglement, but which are nevertheless useless at proving three-particle nonlocality. In particular, it is shown that the quantum predictions giving a maximal violation of Mermin's three-particle Bell inequality [Phys. Rev. Lett. \textbf{65}, 1838 (1990)] can be reproduced by a hybrid hidden variables model in which nonlocal correlations are present only between two of the particles. It should be possible, however, to test the existence of both three-particle entanglement and three-particle nonlocality for any given quantum state via Svetlichny's inequality.
\end{abstract}


\pacs{03.65.Ud,
03.65.Ta}

\maketitle

Referring to pure quantum states, one usually thinks of $n$-particle entanglement and $n$-particle nonlocality (or nonseparability) as equivalent physical notions, in the sense that the quantum-mechanical violation of a $n$-particle Bell inequality (BI) by a $n$-particle entangled state can be explained as a result of the existence of nonlocal quantum correlations relating each of the particles with all the others. While this view is valid for the two-particle case, it is no longer justified for the multiparticle case ($n\geq 3$) since in this case a violation of a $n$-particle BI by a $n$-particle entangled state is \textit{not\/} sufficient for the confirmation of genuine $n$-particle nonlocality. Indeed, one can try to successfully reproduce such $n$-particle BI violation by means of a model in which nonlocal correlations develop only between $m$ ($m<n$) of the particles (where $m$ particles are actually acting nonlocally and can vary from one run of the experiment to the other). This fact was first pointed out by Svetlichny in 1987 \cite{Svetlichny}, who, furthermore, derived an inequality for the three-particle case which is obeyed by such models that assume two-particle nonlocality, but which can be violated by some quantum states thus showing that such states are truly three-particle nonseparable. Svetlichny's inequality (SI) has been generalized to the case of $n$-particle systems in two recent papers \cite{SS02,CGPRS02}.

In this paper we concentrate on the three-particle case and show that Svetlichny's original inequality for distinguishing between two-particle and three-particle nonlocalities is actually also a Bell inequality for three-particle systems. As we will see, this follows immediately from the consideration of the three-particle Bell inequality derived by Mermin \cite{Mermin90a}, and its subsequent comparison with Svetlichny's. Clearly, any three-particle nonseparable state (that is, a state that violates SI) is also a three-particle entangled state. There are, however, three-particle entangled states that violate a given Bell inequality, but which do not violate Svetlichny's. The most significant example of this occurs for the Greenberger-Horne-Zeilinger (GHZ) state when one tries to exhibit an ``all-or-nothing'' type contradiction between quantum mechanics (QM) and local hidden variables (LHV) theories \cite{GHZ89,GHSZ90+M90b}. The measurements involved in this case can lead to a maximal violation of  Mermin's three-particle BI, but, as we will see, such measurements cannot yield a violation of SI, and therefore, cannot be used to determine the three-particle nonseparability of the GHZ state. In other words, the correlations between the results of measurements performed on an ensemble of triplets of particles in the GHZ state showing a maximal violation of Mermin's inequality admits a description in terms of a model in which only two-particle nonlocal correlations are present. We emphasize, however, that, in principle, it is always possible to unambiguously determine the existence of both three-partite entanglement and genuine three-party nonlocality for any given three-particle quantum state. This can be done by experimentally testing Svetlichny's inequality for different combinations of the measurement settings. If a violation of SI does occur for a given set of measurements, then the concurrence of the three-particle entanglement feature and the three-particle nonlocality of the given quantum state is demonstrated. On the other hand, if no violation of SI is observed for any set of measurements, then one can assert that the given quantum state is neither three-particle entangled nor three-particle nonlocal. 

We notice that the the topic treated in this paper, namely the distinction between the concepts of three-particle entanglement and three-particle nonlocality, has already been considered in Refs.\ \cite{SS02,CGPRS02,MPR02,SU02,U02,GH98} (besides Ref.~\cite{Svetlichny}). Specifically, the works in Refs.\ \cite{SS02,CGPRS02,SU02,U02,GH98} show the necessity for experiments aimed at demonstrating $n$-particle nonlocality ($n\geq 3$) to be analyzed using generalized Svetlichny inequalities, while the work in Ref.\ \cite{MPR02} reaches this conclusion for the particular case $n=3$ (see also Ref.\ \cite{CGPRS02} for a detailed study of the case $n=3$). Similarly, in this paper we show that Mermin's three-particle Bell inequality is not adequate to detect full three-particle nonseparability, and conclude that, in order to achieve this goal, one should instead consider Svetlichny's original inequality. We believe, however, that the arguments presented in this paper are, on the one hand, more simple and direct than those in Ref.\ \cite{CGPRS02}, and, on the other hand, more general than those in Ref.\ \cite{MPR02}, and it is our hope that they will lead to a better understanding of the subject discussed.

Consider a situation in which triplets of particles in a (possibly unknown) pure quantum state are emitted by a source. We can regard the three particles in any given triplet as flying apart from the source, so that each of the particles subsequently enters its own measuring station where, for each run of the experiment, one of two possible alternative measurements is performed: $A$ or $A^{\prime}$ on particle 1, $B$ or $B^{\prime}$ on particle 2, and $C$ or $C^{\prime}$ on particle 3. Each of the measurements gives the outcomes to be either $+1$ or $-1$. The basic entity to be considered is the correlation function $E(ABC)$ that represents the expectation value of the product of the measurement outcomes of the observables $A$, $B$, and $C$. Thus Svetlichny's inequality can be written in the form \cite{Svetlichny,MPR02}
\begin{align}
|S_{\text{V}}| = |& E(ABC)+E(ABC^{\prime})+E(AB^{\prime}C)+E(A^{\prime}BC) \nonumber  \\
& - E(A^{\prime}B^{\prime}C^{\prime})-E(A^{\prime}B^{\prime}C)
-E(A^{\prime}BC^{\prime}) \nonumber \\
& -E(AB^{\prime}C^{\prime})| \leq 4 .
\label{SI}
\end{align}
In Ref.\ \cite{MPR02} (see also Refs.\ \cite{SS02,CGPRS02}), it has been shown that the combination of \textit{quantum\/} correlations appearing in inequality (\ref{SI}) is bound by $|S_{\text{V}}|\leq 4\sqrt{2}$, with the maximum quantum violation being attained by the GHZ state for a suitable choice of measurements $A$, $A^{\prime}$, $B$, $B^{\prime}$, $C$, and  $C^{\prime}$.

Let us now recall the Bell-type inequality derived by Mermin for three spin-1/2 particles \cite{Mermin90a}. This inequality imposes an upper limit on the absolute value of a combination of \textit{four\/} correlation functions, which must be satisfied by any LHV theory. One possible form of Mermin's inequality is
\begin{align}
|M| = | & E(ABC^{\prime})+E(AB^{\prime}C)+ E(A^{\prime}BC)  \nonumber  \\
& - E(A^{\prime}B^{\prime}C^{\prime})| \leq 2 \, .
\label{M1}
\end{align}
Clearly, by renaming the observables so that the primed ones $A^{\prime}$, $B^{\prime}$, and $C^{\prime}$ become, respectively, the unprimed ones $A$, $B$, and $C$, and vice versa, one can equally express Mermin's inequality as
\begin{align}
|M^{\prime}| = |& E(ABC)-E(AB^{\prime}C^{\prime})-E(A^{\prime}BC^{\prime}) \nonumber  \\
& - E(A^{\prime}B^{\prime}C)| \leq 2 \, .
\label{M2}
\end{align}
Then, as $S_{\text{V}}=M+M^{\prime}$, we have that $|S_{\text{V}}| = |M+M^{\prime}| \leq |M|+|M^{\prime}| \leq 4$, where the last inequality follows from Eqs.\ (\ref{M1}) and (\ref{M2}). Thus, from the Bell inequalities (\ref{M1}) and (\ref{M2}), a third one has been derived, $|S_{\text{V}}| \leq 4$, which is formally identical to inequality (\ref{SI}). Svetlichny's original inequality can therefore also be interpreted as a Bell inequality and, as such, must be violated by any three-particle entangled state for appropriate choices of observables.

For three spin-1/2 particles in the state $|GHZ\rangle = \tfrac{1}{\sqrt{2}}(|\uparrow\uparrow\uparrow\rangle + |\downarrow\downarrow\downarrow\rangle)$ [with $\uparrow$ ($\downarrow$) denoting spins polarized ``up'' (``down'') along the $z$ axis], quantum mechanics predicts
\begin{align*}
E_{\text{GHZ}}(ABC) &=  \langle GHZ| 
\sigma({\hat{\textbf {n}}}_{1})\otimes \sigma({\hat{\textbf {n}}}_{2}) \otimes 
\sigma({\hat{\textbf {n}}}_{3}) |GHZ \rangle  \\
& = \cos (\phi_1+\phi_2+\phi_3), 
\end{align*}
where, for the sake of simplicity, we have restricted our attention to spin measurements along directions lying in the $x$-$y$ plane, so that such directions ${\hat{\textbf {n}}}_i$ and ${\hat{\textbf {n}}}_{i}^{\prime}$ are specified by the azimuthal angles $\phi_i$ and $\phi_i^{\prime}$, respectively, for $i=1,2,3$. For the choice of angles $\phi_1+\phi_2+\phi_3 =n\pi$ ($n=0,\pm 1,\pm 2,\ldots\,$) and $\phi_i^{\prime}=\phi_i +\tfrac{\pi}{2}$, quantum mechanics gives $|M^{\prime}| =4$, $|M| =0$, and $|S_{\text{V}}|= 4$. Alternatively, for the choice of angles $\phi_1+\phi_2+\phi_3 =(n+\tfrac{1}{2})\pi$ and $\phi_i^{\prime}=\phi_i +\tfrac{\pi}{2}$, quantum mechanics gives $|M| =4$, $|M^{\prime}| =0$, and $|S_{\text{V}}|= 4$. So we can see that the measurements giving a maximal violation of Mermin's inequality do not violate Svetlichny's, and then they do not serve to verify the three-particle nonlocality of the GHZ state. As another example of a quantum violation of a Bell inequality, but not of Svetlichny's, we may take the angles $\phi_1+\phi_2+\phi_3 =\tfrac{\pi}{6}$ and $\phi_i^{\prime}=\phi_i +\tfrac{\pi}{2}$. For this case quantum mechanics predicts $M^{\prime} = 3.46$ and $S_{\text{V}}= 1.46$. On the other hand, the maximal violation of SI is attained for the values $\phi_1+\phi_2+\phi_3 =(n+\tfrac{3}{4})\pi$ and $\phi_i^{\prime}=\phi_i +\tfrac{\pi}{2}$. These give $|M| =|M^{\prime}|=2\sqrt{2}$ and $|S_{\text{V}}|= 4\sqrt{2}$ \footnote{%
In Ref.\ \cite{MPR02}, it has been shown that, in order for SI to be maximally violated by a quantum state $|\psi\rangle$, it is necessary that $\langle\psi|CC^{\prime}+C^{\prime}C|\psi\rangle =0$. For a spin-1/2 particle, this condition is met whenever the measurement directions corresponding to the spin observables $C$ and $C^{\prime}$ are perpendicular between themseleves, since this is equivalent to the vanishing of the anticommutator $\{C,C^{\prime}\}$. By symmetry, the conditions $\langle\psi|AA^{\prime}+ A^{\prime}A|\psi\rangle =\langle\psi|BB^{\prime}+B^{\prime}B|\psi\rangle=0$ must also be met if the state $|\psi\rangle$ is to maximally violate SI.}.

In view of the results in the preceding paragraph, it will be argued that, actually, the three-particle nonseparability feature of a (three-particle) quantum state cannot be tested on the basis of only four correlation functions pertaining to either the set $E_1 \equiv \{E(ABC^{\prime}),E(AB^{\prime}C),E(A^{\prime}BC),E(A^{\prime} B^{\prime}C^{\prime})\}$ or the set $E_2 \equiv \{E(ABC),E(AB^{\prime}C^{\prime}),E(A^{\prime}BC^{\prime}) ,E(A^{\prime}B^{\prime}C)\}$. Indeed, it is important to realize that one can always reproduce whatever values are assumed by four such functions in either $E_1$ or $E_2$ by means of a hidden variables model in which arbitrary (nonlocal) correlations are permitted between two of the particles, but only local correlations between these two particles and the third one. As an example illustrating this point, consider the case in that we are given the values: $E(ABC)=+1$ and $E(AB^{\prime}C^{\prime})=E(A^{\prime}BC^{\prime}) =E(A^{\prime}B^{\prime}C)=-1$. These perfect correlations maximally violate Mermin's inequality (\ref{M2}), but they are nevertheless consistent with a hybrid local-nonlocal hidden variables model \cite{Svetlichny,CGPRS02,MPR02} in which, for example, particles 1 and 2 form a nonlocal subsystem, and this subsystem is locally correlated with particle 3. The simplest model of this kind one can think of is one for which the outcomes of the nonlocal measurements $AB$, $AB^{\prime}$, $A^{\prime}B$, and $A^{\prime}B^{\prime}$, as well as the outcomes of the local ones $C$ and $C^{\prime}$, are completely determined (with probability either $0$ or $1$) by the value of a hidden variable $\lambda$. Note that the hidden variable is \textit{not\/} allowed to determine the outcomes of the local measurements $A$, $A^{\prime}$, $B$, and $B^{\prime}$, since, for the considered model, particles 1 and 2 are assumed to be nonlocally correlated. With this in mind, one can conceive a trivial hybrid hidden variables model which, for each value of $\lambda$, yields the outcomes $AB=+1$, $AB^{\prime}=+1$, $A^{\prime}B=+1$, $A^{\prime}B^{\prime}=-1$, $C=+1$, and $C^{\prime}=-1$. This particular model then gives $M^{\prime} =4$, $M=0$, and $S_{\text{V}}= 4$. On the other hand, one could devise a more elaborate hybrid local-nonlocal model that reproduces the set of quantum correlations $C_1 \equiv \{E(ABC^{\prime})= E(AB^{\prime}C)= E(A^{\prime}BC) =-E(A^{\prime}B^{\prime}C^{\prime})=\tfrac{1}{\sqrt{2}} \}$. Likewise, one could devise \textit{another\/} hybrid model that reproduces the set of quantum correlations $C_2 \equiv \{E(ABC)= -E(AB^{\prime}C^{\prime})= -E(A^{\prime}BC^{\prime}) =-E(A^{\prime}B^{\prime}C)=\tfrac{1}{\sqrt{2}} \}$. However, no hybrid model exists that accounts simultaneously for the full set of correlations $C_1 \cup C_2$. This indicates that, in order to verify the existence of genuine three-particle nonlocality for a given quantum state, it is necessary to consider the expectation value of all eight product observables $ABC,ABC^{\prime},\ldots,A^{\prime}B^{\prime}C^{\prime}$. If these values violate SI for some choice of $A$, $A^{\prime}$, $B$, $B^{\prime}$, $C$, and  $C^{\prime}$, then the three-particle nonseparability of the given quantum state would have been unambiguously demonstrated.

It is to be noted, on the other hand, that Mermin's inequality does discriminate between the three-particle states that are three-particle entangled, and the three-particle states that are not. In fact, it is known \cite{GH98} that if the value of either $|M|$ or $|M^{\prime}|$ predicted by quantum mechanics for a given three-particle state $|\psi\rangle$ is greater than $2\sqrt{2}$, then $|\psi\rangle$ exhibits true three-particle entanglement. Otherwise, $|\psi\rangle$ is at most two-particle entangled. The arguments given in the preceding paragraph, however, tell us that Mermin's inequality cannot discriminate the purely quantum-mechanical correlation due to the entanglement between all three particles (e.g., in the GHZ state) from such hybrid models that assume nonlocal correlation only between two of the three particles. By contrast, Svetlichny's inequality is capable of detecting both three-particle entanglement and three-particle nonseparability. In fact, whenever the quantum prediction of the quantity $|S_{\text{V}}|$ for a given three-particle state $|\psi\rangle$ is greater than $4$, then $|\psi\rangle$ is three-particle entangled, and the quantum correlations developing between the three particles can definitely not be reproduced by any hybrid model of the kind discussed. In any case, since both Mermin's and Svetlichny's inequalities detect quantum entanglement, one could take the amount of violation (properly normalized) of either one of them as a measure of the amount of entanglement present in the state $|\psi\rangle$.

To conclude, we briefly discuss some of the predictions that quantum mechanics makes for the so-called W state, $|W\rangle = \tfrac{1}{\sqrt{3}} (|\uparrow\uparrow\downarrow\rangle + |\uparrow\downarrow\uparrow\rangle + |\downarrow\uparrow\uparrow\rangle)$. The consideration of this state is important by itself since, as was shown in Ref.\ \cite{DVC00}, any nontrivial three-particle entangled state can be converted, by means of invertible local operations and classical communication (ILOCC), into one of the two inequivalent forms (under ILOCC) of genuine tripartite entanglement whose representative states appear to be the GHZ state and the $W$ state. When we restrict to spin measurements in the $x$-$z$ plane, the expectation value of the product observable $\sigma(\theta_{1})\otimes \sigma(\theta_{2}) \otimes \sigma(\theta_{3})$ predicted by quantum mechanics for the $W$ state is \footnote{
It can be seen, on the other hand, that, for spin measurements in the $x$-$y$ plane, the predicted quantum expectation value for the $W$ state vanishes for any choice of azimuthal angles, $E_{\text{W}}(\phi_1,\phi_2,\phi_3)=0$.}
\begin{align}
E_{\text{W}}(\theta_1,\theta_2,\theta_3) = & -\frac{2}{3} \cos (\theta_1 +\theta_2 +\theta_3) \nonumber  \\
& - \frac{1}{3} \cos\theta_1 \cos\theta_2 \cos\theta_3  ,
\end{align}
where the polar angle $\theta_1$ ($\theta_{1}^{\prime}$) specifies the measurement direction of the spin observable $A$ ($A^{\prime}$), etc. For the choice of angles $\theta_1=\theta_2=\theta_3=\tfrac{1}{3}n\pi$ ($n=0,\pm 1,\pm 2,\ldots\,$) and $\theta_i^{\prime}=\theta_i +\tfrac{\pi}{2}$ quantum mechanics gives $|M^{\prime}| =3$, $|M| =0$, and $|S_{\text{V}}|= 3$. Alternatively, for the choice of angles $\theta_1=\theta_2=\theta_3=(\tfrac{1}{6}+\tfrac{1}{3}n)\pi$ and $\theta_i^{\prime}=\theta_i +\tfrac{\pi}{2}$, quantum mechanics gives $|M| =3$, $|M^{\prime}|=0$, and $|S_{\text{V}}|= 3$. For measurement directions fulfilling the conditions $\theta_1=\theta_2=\theta_3$ and $\theta_{1}^{\prime}= \theta_{2}^{\prime}= \theta_{3}^{\prime}$, the maximum value of either $|M^{\prime}|$ or $|M|$ that can be obtained for the $W$ state is 3.046 \cite{Cabello02}. This maximum value occurs, for instance, for the angles $\theta_i=54.032^{\circ}$ and $\theta_{i}^{\prime}=156.106^{\circ}$, which gives $M^{\prime}=3.046$, $M=0.054$, and $S_{\text{V}}=3.1$. For all these sets of directions the $W$ state violates Mermin's inequality, but not Svetlichny's. On the other hand, considering again spin observables for which $\theta_1=\theta_2=\theta_3$ and $\theta_{1}^{\prime}= \theta_{2}^{\prime}= \theta_{3}^{\prime}$, the maximal violation of SI predicted by quantum mechanics for the $W$ state is found to be $|S_{\text{V}}|=4.354$ (with $|M| =|M^{\prime}|=2.177$), which is obtained, for instance, for the choice of angles $\theta_i=35.264^{\circ}$ and $\theta_{i}^{\prime}=144.736^{\circ}$. We note that the maximum value of $|S_{\text{V}}|$ attained by the GHZ state is greater than that attained by the $W$ state. Loosely speaking, this means that the three particles in the GHZ state are more strongly correlated than they happen to be when coupled in the $W$ state. As a counterpart, however, W is the three-qubit state that maximally retains bipartite entanglement when any one of the three qubits is traced out \cite{DVC00}. Indeed, it is easily seen that when any one of the particles in the GHZ state is measured in the basis $\{\uparrow, \downarrow\}$ then the other two are invariably left in a product state. On the other hand, when any one of the particles in the $W$ state is measured in the same basis then there exists a probability $2/3$ that the other two particles are left in a maximally entangled state.

In summary, we have shown that there are sets of measurements, which produce a violation of Memin's three-particle Bell inequality, but which do not allow Svetlichny's inequality to be violated. Hence such measurements cannot be used for the verification of genuine three-particle nonlocality, although they can be used to demonstrate three-particle quantum entanglement. Examples of such measurements have been given for three spin-1/2 particles in either the GHZ state or the $W$ state. In particular, we have shown that the quantum correlations leading to an ``all or nothing'' type contradiction between QM and LHV theories, can be described by a model in which nonlocal correlations take place only between two of the particles. Moreover, we have argued that any four expectation values in either the set $E_1$ or $E_2$ can be reproduced by a hybrid local--two-particle-nonlocal hidden variables model. It is therefore necessary to consider more than four correlation functions if we want to distinguish between two-particle and three-particle nonlocality. This can be achieved via Svetlichny's inequality. Finally, we note that an open question left is what is the \textit{minimum\/} number of correlation functions one should consider in order to discriminate between two-particle and three-particle nonlocality.

The author wishes to thank Ad\'{a}n Cabello for useful comments and discussions.

\end{document}